\documentclass[aps,pre,twocolumn,groupedaddress,showpacs]{revtex4}

\usepackage{graphicx}

\usepackage{hyperref}

\usepackage{color}

\begin{document}


\title{Discontinuous nonequilibrium phase transitions in a nonlinearly
  pulse-coupled excitable lattice model}


\author{Vladimir R. V. Assis}
\email{vladimirassis@df.ufpe.br}
\thanks{corresponding author}
\author{Mauro Copelli}%
\email{mcopelli@df.ufpe.br}
\affiliation{%
Laborat\'orio de F{\'\i}sica Te\'orica e Computacional, Departamento
de F{\'\i}sica, Universidade Federal de Pernambuco, 50670-901 Recife, PE, Brazil}%

\date{\today}

\begin{abstract}
We study a modified version of the stochastic
susceptible-infected-refractory-susceptible (SIRS) model by employing
a nonlinear (exponential) reinforcement in the contagion rate and no
diffusion. We run simulations for complete and random graphs as well
as $d$-dimensional hypercubic lattices (for $d=3,2,1$). For weak
nonlinearity, a continuous nonequilibrium phase transition between an
absorbing and an active phase is obtained, such as in the usual
stochastic SIRS model [Joo and Lebowitz, Phys. Rev. E {\bf 70}, 036114
  (2004)]. However, for strong nonlinearity, the nonequilibrium
transition between the two phases can be discontinuous for $d\geq 2$,
which is confirmed by well-characterized hysteresis cycles and
bistability.  Analytical mean-field results correctly predict the
overall structure of the phase diagram. Furthermore, contrary to what
was observed in a model of phase-coupled stochastic oscillators with a
similar nonlinearity in the coupling [Wood {\it et al.\/},
  Phys. Rev. Lett. {\bf 96}, 145701 (2006)], we did not find a
transition to a stable (partially) synchronized state in our
nonlinearly pulse-coupled excitable elements. For long enough
refractory times and high enough nonlinearity, however, the system can
exhibit collective excitability and unstable stochastic oscillations.
\end{abstract}










\maketitle

\section{Introduction}

Understanding collective effects of noisy excitable elements is
essential for several disciplines, such as neuroscience, epidemiology,
and chemistry, among others. An isolated excitable element is a
dynamical system which stays in a quiescent state until it suffers a
sufficiently strong perturbation.  In that case its trajectory in
phase space can be characterized by an excited state, which is then
followed by a refractoriness to further perturbations before returning
to rest. A minimal (discrete) model of an excitable element therefore
consists of a three-state system~\cite{Lindner04}. In the parlance of
neuroscience (epidemics), each element could represent a neuron or
patch of active membrane (individual) which sequentially becomes
polarized (susceptible), spiking or depolarized (infected), and then
refractory (recovered). Collective dynamics emerge because quiescent
elements are typically perturbed by excited elements.

A simple system which incorporates these ingredients in a scenario
with noise is the stochastic
susceptible-infected-recovered-susceptible (SIRS) epidemic lattice
model~\cite{Joo04b}. It is a continuous-time model in which a site
goes from susceptible to infected at a rate which depends on the
density of its infected neighbors. In epidemiology, the model (or
variants thereof) can be employed to investigate, e.g., whether an
initial density of infected sites (which is usually chosen as the
order parameter) will reach a nonzero stationary value or decrease to
zero. If all sites become quiescent, the dynamic halts and the system
is said to be in a absorbing state~\cite{Marro99}.

Increasing the coupling between infected and susceptible sites, the
SIRS model undergoes a continuous nonequilibrium phase transition from
an absorbing to an active phase characterized by a stationary nonzero
density of infected sites. However, experimental data (from
neuroscience, epidemiology, and chemistry, among others) can exhibit
also global oscillations. This additional transition has been observed
mostly in cellular automata, where the sites are synchronously
updated~\cite{lewis00,Kuperman01,Lindner04,Kinouchi06a}. This
technical detail is apparently relevant since global oscillations in
stochastic continuous-time models are less common. As discussed by
Risau-Gusman and Abramson, they sometimes appear as stochastic
oscillations in single runs of the model but disappear in
trajectories of the averaged lattice activity and analytical
descriptions (usually mean-field)
thereof~\cite{Risau-Gusman07}. Global oscillations predicted by
mean-field (MF) approximations were observed in both
non-Markovian~\cite{Prager03} and Markovian~\cite{Wood06a} models of
three-state continuous-time stochastic oscillators. These models,
however, do not have an absorbing state.

\begin{figure*}[!t]
\includegraphics[width=1.\textwidth,angle=0]{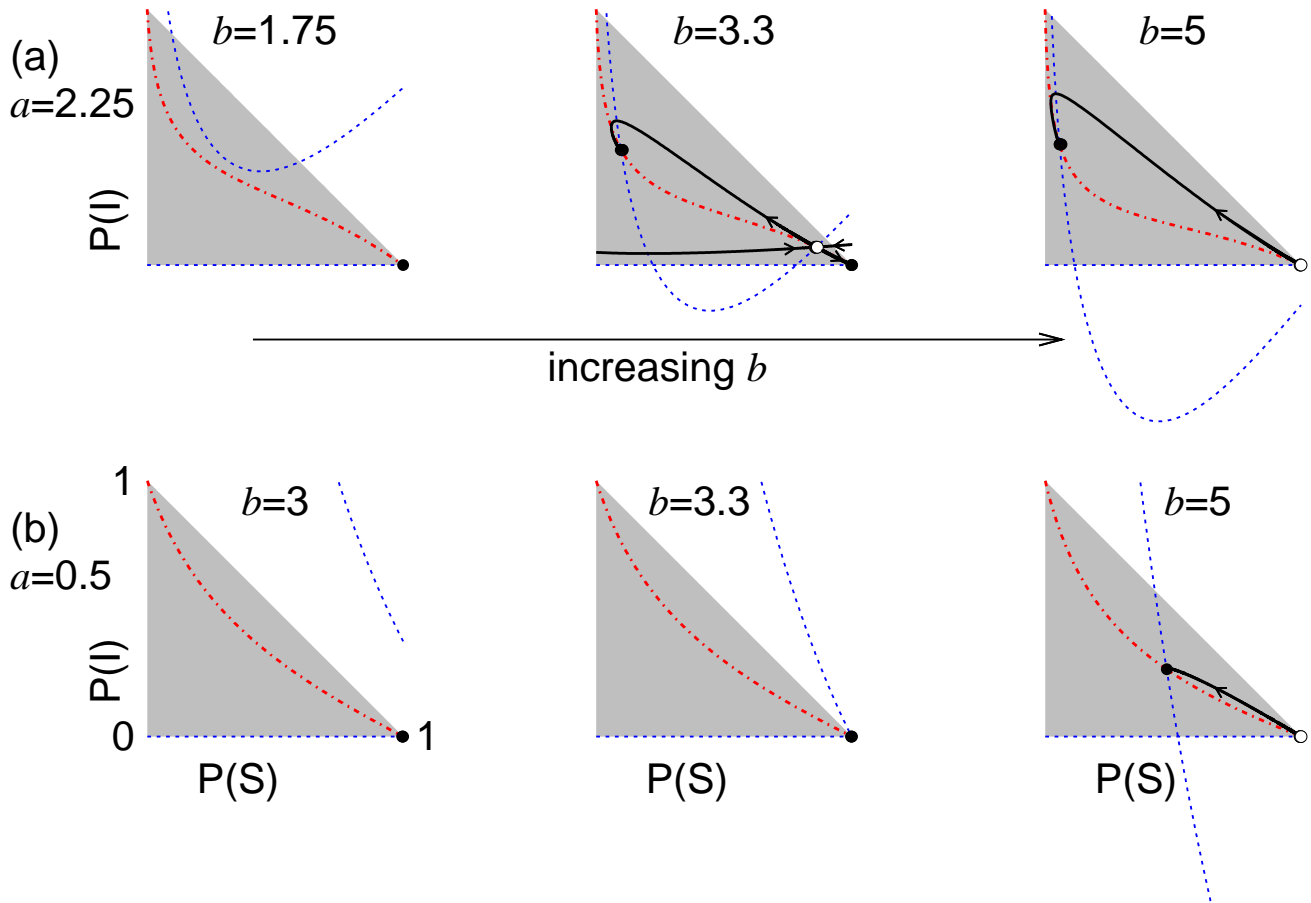}%
\begin{center}
\caption{\label{fig:phasespace}(Color online) Phase space $[P(S),
  P(I)]$ showing the nullclines $\dot P(I) =0$ (dashed lines) and
  $\dot P(S) =0$ (dot-dashed lines). Grey triangles corresponds to the
  physically acceptable region $0 \leq P(I) \leq 1-P(S)$. In each row,
  $a$ is kept constant and $b$ increases from left to right. In the
  lower panel (b) ($a=0.5$), the absorbing state at $P(S)=1$ loses
  stability in a transcritical bifurcation as $b$ increases. In the
  upper panel (a) ($a=2.25$), the $\dot P(I) =0$ nullcline is bent by
  the exponential nonlinearity [see eq.~\ref{eq:ab}], creating an
  active state with finite $P(I)$ in a saddle-node bifurcation. Closed
  (open) symbols denote stable (unstable) fixed points. Note that the
  nullcline $\dot P(I)=0$ always comprises the line $P(I)=0$. When the
  absorbing state is unstable (a saddle), $P(I)=0$ corresponds to its
  stable manifold (rightmost column). Except for this case, the
  trajectories (solid lines) represent the stable and unstable
  manifolds of the saddles.}
\end{center}
\end{figure*}

Here we investigate a modified version of the SIRS model with a
nonlinear rate which can be considered a {\em pulse-coupled excitable}
version of the original {\em phase-coupled oscillator} model of Wood
{\it et al.\/}~\cite{Wood06a}. We will show, on the one hand, that
this nonlinear (but Markovian) extension of the SIRS model is
apparently insufficient for the generation of sustained collective
oscillations. On the other hand, the model presents phase transitions
into an absorbing state which can be continuous (if weakly nonlinear)
or discontinuous (if nonlinear enough and for $d\geq 2$). We therefore
provide a three-state continuous-time model which can undergo
discontinuous phase transitions like those of Bidaux {\it et al.\/}
for a two-state cellular automaton~\cite{Bidaux89} except that the
nature of the transition here can be controlled by a free parameter
(not only the spatial dimension~\cite{Bidaux89}). Moreover,
differently from most models presenting nonequilibrium discontinuous
phase transitions~\cite{DickmanTome91,OdorBoccaraSzabo93,Marro99}, the
parameter which controls the nature of the transition is not diffusion
(which is absent from our model). The model can also exhibit a
different bifurcation scenario from what is usually observed in
nonequilibrium lattice models, showing collective excitability and
unstable global oscillations.

The paper is organized as follows. In Sec.~\ref{model} we introduce
the model, analyze its mean-field solution, and compare it to
simulations of random and complete graphs. Simulation results for
$d$-dimensional hypercubic lattices are shown in
Sec.~\ref{simulations}, while our concluding remarks are discussed
in Sec.~\ref{conclusions}.

\section{\label{model}Model}

In the conventional stochastic version of the SIRS epidemic model, a
susceptible ($S$) site at position $x$ ($x=1,\ldots,N$) becomes
infected ($I$) at a rate $\lambda n_I(x)/k(x)$, where $k(x)$ is the
number of neighbors of $x$, out of which $n_I(x)$ are infected, and
$\lambda$ is the so-called infection rate (the coupling
parameter). After that, the site becomes temporarily insensitive to
its surroundings (hence the term pulse coupling), jumping from
infected to recovered ($R$) at a constant rate $\delta$, and from
recovered back to susceptible at a rate $\gamma$. This is summarized
as follows:
\begin{eqnarray}
\label{eq:lambda}S\longrightarrow I && \mbox{at rate } \lambda n_I/k, \\ \nonumber \\
\label{eq:delta}I\longrightarrow R && \mbox{at rate } \delta, \\ \nonumber \\
\label{eq:gamma}R\longrightarrow S && \mbox{at rate } \gamma.
\end{eqnarray}
Since all rates could in principle be normalized by $\delta$, in the
following we set $\delta=1$ without loss of generality.

For low values of $\lambda$, an initial density of infected sites
eventually dies out and the system reaches its unique absorbing state
(all sites susceptible). For $\lambda$ larger than a critical value
$\lambda_c$, on the other hand, a phase with nonzero density of
infected sites becomes stable in the thermodynamic limit $N\to\infty$
(though the absorbing state is obviously always a solution). The
transition at $\lambda_c$ (studied in detail by Joo and
Lebowitz~\cite{Joo04b}) is widely believed to be continuous and
belonging to the directed percolation (DP) universality
class~\cite{Assis08,deSouza09}.

Adapting the nonlinear coupling employed by Wood {\it et
  al.}~\cite{Wood06a} to pulse-coupled excitable elements, one obtains
a generalization of the SIRS model. Instead of Eq.~\ref{eq:lambda}, we
propose
\begin{eqnarray}
& S\longrightarrow I \mbox{ at rate} & \nonumber \\ \nonumber \\
\label{eq:ab}
& g\left(n_I/k,n_S/k\right) \equiv b \left[ e^{a(n_I-n_S)/k} - e^{-an_S / k} \right], &
\end{eqnarray}
where $a$ and $b$ are coupling parameters, $n_S$ is the number of
susceptible neighbors, and Eqs.~\ref{eq:delta} and~\ref{eq:gamma}
remain unaltered. Note that the interaction occurs only among first
neighbors.  The second term in rate~\ref{eq:ab} guarantees the
existence of an absorbing state: if all sites are susceptible (thus
$n_I = 0$ for all sites), they will remain susceptible forever. For
small values of $a$ ($a\ll 1$), one recovers the linear behavior of
the original SIRS model, with $\lambda \simeq ab$, to first order. So
for small $a$, increasing $b$ leads to a continuous phase transition
just like in the SIRS model~\cite{Joo04b}. For large enough $a$,
however, we will show that increasing $b$ leads to a discontinuous
phase transition.

As a motivation to rate~\ref{eq:ab}, let us note that the dynamics
underlying neuronal firing is highly nonlinear in several of its
aspects: membrane depolarizes when (typically Na$^+$) channels open
very quickly, in a thresholdlike behavior. This process, on its turn,
is triggered by a (nonlinear) sum of smaller depolarizations induced
at synapses by presynaptic neurons. Synaptic dynamics (including
neurotransmitter binding, in the simplest case) is itself
nonlinear. It is therefore not unusual that reduced models of
collective neuronal phenomena allow for nonlinear
terms~\cite{Gerstner}. In our case, the nonlinearity is controlled by
parameter $a$.

\subsection{Mean-field analysis}

The structure of the phase diagram in the $(a,b)$ plane can be
captured most easily by analyzing the mean-field version of the
model. Letting $P(\alpha)$ be the probability that a site is in state
$\alpha$ ($\alpha \in \{S,I,R \}$), one obtains
\begin{eqnarray}
\label{eq:ps}
\dot P(S) & = & - g[P(I),P(S)]P(S) + \gamma P(R), \\ \nonumber \\
\label{eq:pi}
\dot P(I) & = & g[P(I),P(S)]P(S) - \delta P(I), \\ \nonumber \\
\label{eq:pr}
\dot P(R) & = & \delta P(I) -\gamma P(R).
\end{eqnarray}
These equations are exact for a complete graph when $k = N\to \infty$
and $n_{\alpha}\to\infty$ with $n_{\alpha}/k=P(\alpha)$. In what
follows, we will employ the stationary value of the density of
infected sites $P(I)^*$ as the order parameter.

By employing the normalization condition $P(S)+P(I)+P(R)=1$, one can
eliminate $P(R)$ and study the resulting two-dimensional flow of the
mean-field dynamics in the $[P(S), P(I)]$ plane, as shown in
Fig.~\ref{fig:phasespace}. For low values of $a$ (lower panel of
Fig.~\ref{fig:phasespace}), the absorbing state $P(S)=1$ loses
stability in a transcritical bifurcation, giving rise to a stable
active state with a density of infected sites which increases
continuously from $P(I)=0$. For large enough $a$, the discontinuous
character of the phase transition reveals itself in the mean-field
equations through a saddle-node bifurcation (upper panel of
Fig.~\ref{fig:phasespace}): an active state appears with nonzero
$P(I)$, while the absorbing state remains stable. As usual, this
bistability is regulated by the stable manifold of the saddle, which
separates the basins of attraction of the two stable fixed
points. Increasing $b$ further, another transcritical bifurcation
occurs: the absorbing state loses stability and the active state
becomes the only attractor of the system.

\begin{figure}[!tb]
\includegraphics[width=0.95\columnwidth,angle=0]{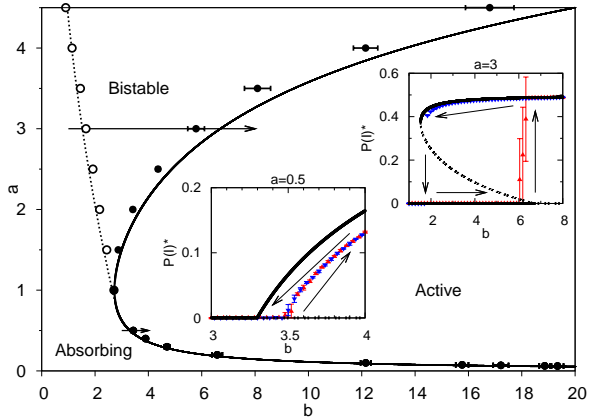}%
\begin{center}
\caption{\label{fig:phasediagram}(Color online) Phase diagram of the
  (mean-field version of the) model. The dashed line marks the onset
  of phase bistability, with the sudden appearance of a stable active
  phase with finite order parameter [saddle-node bifurcation in
  Eqs.~\ref{eq:ps}-\ref{eq:pr}]. Solid lines denote a transition in
  which the absorbing state loses stability [transcritical bifurcation
  in Eqs.~\ref{eq:ps}-\ref{eq:pr}]. Symbols correspond to simulations
  in a random graph with $\gamma=1$, $K=10$, $N=10^4$,
  $t_{max}=5\times 10^3$, and $h=10^{-5}$ averaged over five runs. The
  lower (upper) inset shows the order parameter $P(I)^*$ for a cyclic
  change in the coupling parameter $b$, showing a continuous
  (discontinuous) transition for a small (large) value of $a$. Upward
  (downward) triangles: increasing (decreasing) $b$.}
\end{center}
\end{figure}

\subsection{Random graph simulations}

Hysteresis is one of the simplest fingerprints of multistability, and
in this system it can be clearly detected in simulations of Erd{\H
  o}s-R{\'e}nyi random graphs~\cite{Albert02} with finite average
connectivity $K$, with which the mean-field equations show a good
agreement. Simulations were performed for a fixed value of $a$.  For
each value of $b$, we allowed the system to evolve during $t_{max}$
time steps. Each time step $\Delta t$ (corresponding to $N$ random
updates~\cite{Marro99}) was chosen to be $(\delta+\gamma+be^a)^{-1}$
to make sure probabilities are less than one.  Parameter $b$ was
increased or decreased in constant intervals $\delta b$, and the
initial condition of the network for each value corresponded to the
final condition of the preceding case. We applied a small rate $h$
($h<1/N$) to spontaneously excite quiescent sites, thus preventing the
system from getting trapped in the absorbing state by finite-size
fluctuations~\cite{Bidaux89,Takeuchi08}.

The insets of Fig.~\ref{fig:phasediagram} show that a loop in $b$
leads to a hysteresis cycle [as observed for the density of active
sites $P(I)$] for high enough values of $a$ ($a>a_c$). For low values
of $a$, the phase transition is continuous. The boundaries between the
phases in the $(a,b)$ plane (main plot of Fig.~\ref{fig:phasediagram})
can be numerically obtained from the mean-field equations and easily
estimated from the random graph simulations (hysteresis cycles did not
change significantly when $t_{max}$ was doubled; see also
Sec.~\ref{simulations}).

To obtain the phase diagram shown in Fig.~\ref{fig:phasediagram}, we
employed a mean refractory time comparable to the excitation time,
$\gamma=1$. However, we did not find qualitative differences in the
mean-field phase diagram in the limiting case $\gamma\to\infty$, which
suggests that the simulation results we present here will also be
valid for a two-state system.

\subsection{Absence of sustained global oscillations}

\begin{figure}[!tb]
\includegraphics[width=0.95\columnwidth,angle=0]{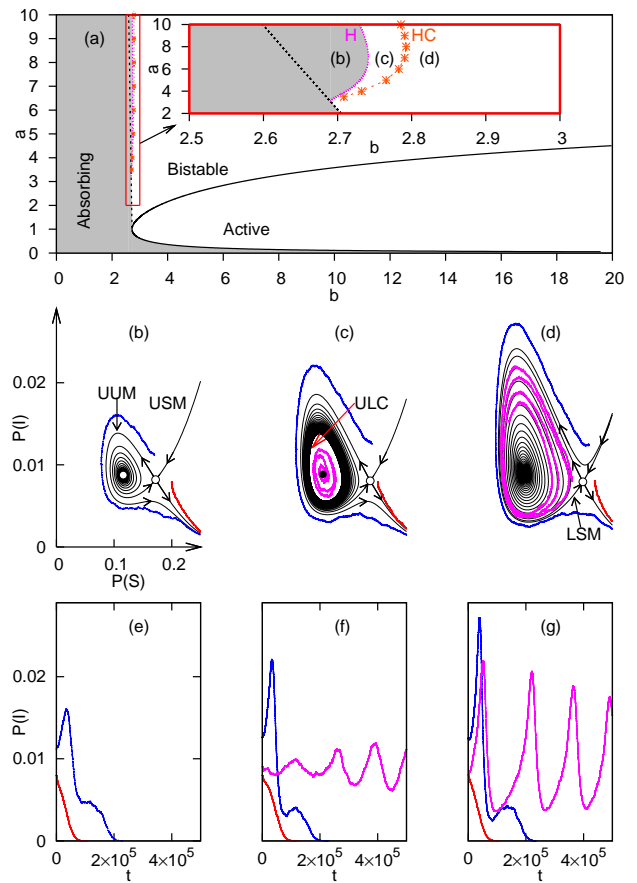}%
\begin{center}
\caption{\label{fig:phasediagram0.01}(Color online) (a) Phase diagram
  of the (mean-field version of the) model for $\gamma=0.01$ (panels
  b--g also show single-run simulations of a complete graph with
  $N=10^6$). Bifurcations are as in Fig.~\ref{fig:phasediagram},
  except for those additionally occurring within the marked rectangle
  (inset): for increasing $b$, a saddle-node bifurcation (dashed line)
  is followed by a Hopf (H) bifurcation (dotted line), after which a
  homoclinic (HC) bifurcation occurs (see text). The absorbing phase
  corresponds to the grey region. The labels (b)--(d) in the inset show
  the points in parameter space which correspond to the phase
  portraits below. (b) A saddle and an unstable fixed point are born
  in a saddle-node bifurcation. (c) The fixed point becomes stable
  via a Hopf bifurcation and is surrounded by an unstable limit
  cycle. (d) After a homoclinic bifurcation, the limit cycle
  disappears. UUM = upper unstable manifold of the saddle; USM = upper
  stable manifold of the saddle; ULC = unstable limit cycle; LSM =
  lower stable manifold. The scale in (b) also applies (apart from a
  translation) to (c) and (d). The lower unstable manifold of the
  saddle always goes to the absorbing fixed point. Panels e, f, and g
  show time series for simulated trajectories, respectively, shown in
  panels b, c, and d, with examples of collective excitability and
  stochastic oscillations (see text for details).}
\end{center}
\end{figure}

In the process of scanning parameter space in search of collective
oscillations, we have found small regions in which a Hopf bifurcation
involving the active state can indeed occur. As we shall see, however,
this does not necessarily imply the existence of sustained collective
oscillations, which we have not found in this model. 

We exemplify with results for $\gamma=0.01$, whose phase diagram is
shown in Fig.~\ref{fig:phasediagram0.01}a. Though qualitatively
similar to Fig.~\ref{fig:phasediagram}, there is now a narrow interval
of $b$ values (for high values of $a$) in which the route to
bistability, instead of the simple saddle-node scenario described in
Fig.~\ref{fig:phasespace}(a), requires additional intermediate
bifurcations~\footnote{For larger values of $\gamma$
  (e.g., $\gamma=1$), the same intermediate bifurcations may occur but
  for larger values of $a$ and restricted to narrower intervals of $b$
  values.} [inset of Fig.~\ref{fig:phasediagram0.01}(a)]. These occur
while the absorbing fixed point remain stable.

Starting from the absorbing phase and increasing $b$, first a
saddle-node bifurcation occurs in which the node is {\em unstable\/}
[and quickly becomes a spiral, Fig.~\ref{fig:phasediagram0.01}(b)]. At
this stage, the system still has only a single stable fixed point, but
the structure of phase space is such that the system has become {\em
  collectively excitable\/}: if the system is below the upper stable
manifold (USM) of the saddle [Fig.~\ref{fig:phasediagram0.01}(b)], it
will monotonously return to the absorbing state, whereas a point above
the USM will go through a long excursion around the upper unstable
manifold (UUM) before coming to rest~[Fig.~\ref{fig:phasediagram0.01}(b)], 
displaying a spike-like time series~[Fig.~\ref{fig:phasediagram0.01}(e)].

Increasing $b$ further, the spiral becomes {\em stable\/} in a Hopf
bifurcation and is now surrounded by an {\em unstable\/} limit cycle
(ULC) [Fig.~\ref{fig:phasediagram0.01}(c)]. Formally, in this parameter
region the system is collectively bistable, but note that the active
phase will only be reached from initial conditions within the ULC
[which, owing to the small value of $\gamma$, is also very small ---
see scale in Fig.~\ref{fig:phasediagram0.01}(b)]. Note also that the
overcrowding of lines outside the ULC signals that it is weakly
repulsive. The inner stable fixed point (the active phase) is
correspondingly weakly attractive (which is the reason why the flux
inside the ULC is not shown). This means that a collective oscillation
solution exists (the ULC) but is so weakly unstable that it might get
confounded with sustained oscillations [even in the numerical
integration of eqs.~\ref{eq:ps}-\ref{eq:pr}]. From single-run
simulations of a complete graph with $N=10^6$, we have obtained a time
series with oscillations [Fig.~\ref{fig:phasediagram0.01}(f)] which only
disappear when, owing to finite-size fluctuations, the system reaches
the absorbing state~\cite{Marro99}.  We therefore confirm the scenario
predicted by Risau-Gusman and Abramson: since the eigenvalues of the
stable fixed point have a nonzero imaginary component, inevitable
fluctuations will generate stochastic
oscillations~\cite{Risau-Gusman07}.

Finally, the ULC disappears in a homoclinic (HC) bifurcation, after which
the active fixed point is separated from the absorbing fixed point
only by the stable manifolds of the saddle. Note that the lower stable
manifold [LSM, see Fig.~\ref{fig:phasediagram0.01}(d)] no longer comes
from the unstable fixed point nor from the ULC but rather joins the
USM as $t\to -\infty$. The LSM will gradually unfold as $b$ increases
until a phase portrait similar to that of the central plot of
Fig.~\ref{fig:phasespace}a is reached (before the saddle collides --
for yet larger values of $b$ -- with the absorbing fixed point in a
transcritical bifurcation). Collective excitability and stochastic
oscillations remain present in this 
regime~[Figs.~\ref{fig:phasediagram0.01}(d) and~\ref{fig:phasediagram0.01}(g)].

Our phase diagram emerged essentially from local stability analysis,
so in principle it does not exclude a saddle-node bifurcation of
cycles from occurring within the active or bistable regions. However,
numerical integration of Eqs.~\ref{eq:ps}-\ref{eq:pr} for a variety of
initial conditions and combination of parameters did not show any
signs of it.

Although the above analysis is based on the mean-field approximation,
it is worth mentioning that improvements on the mean-field
approximation do not necessarily help the prediction of collective
oscillations. Rozhnova and Nunes~\cite{Rozhnova09} recently observed
that the equations obtained by Joo and Lebowitz~\cite{Joo04b} for the
two-site approximation lead to sustained oscillations in a small
region for $\gamma \ll 1$. However, when simulating random graphs in
the same parameter region, these oscillations get
damped~\cite{Rozhnova09b}. 

\section{\label{simulations}Simulations in hypercubic lattices}

\begin{figure}
\includegraphics[width=0.95\columnwidth,angle=0]{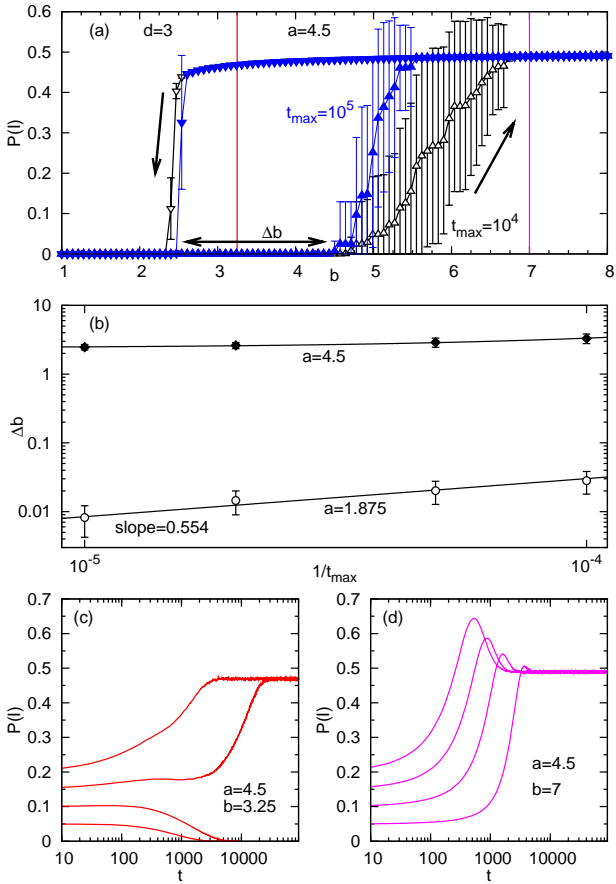}%
\begin{center}
\caption{\label{fig:3d}(Color online) Simulations for $d=3$ with
  $N=25^3$ and $h=2.5 \times 10^{-5}$. (a) ($a=4.5$, averaged over 20
  runs) The length of the hysteresis cycle $\Delta b$ decreases for
  increasing $t_{max}$ ($\Delta b$ denoted by the horizontal arrow for
  $t_{max}=10^5$). Upward (downward) triangles denote increasing
  (decreasing) $b$. Vertical lines show values of $b$ employed in
  panels (c) and (d). (b) Dependence of $\Delta b$ on $t_{max}$ is 
  qualitatively different for $a$ above (filled circles) and below
  (open circles) $a_c$, respectively, showing a nonzero or zero
  asymptotic value in the limit $t_{max}\to\infty$. The lower plot is
  well fitted by a power law with exponent 0.55. Note that the
  precision in $\Delta b$ is limited by the increment in $b$ employed
  in the hysteresis cycle ($\delta b=1.8\times 10^{-3}$ in the
  case $a=1.875$).  Panels (c) and (d) show the time evolution 
  of $P(I)$ (averaged over 20 runs) for $a=4.5$ and different initial
  conditions [with $N P(I)$ sites in state $I$ at $t=0$ and the
  remaining in state $S$], showing phase bistability for $b=3.25$ but
  not for $b=7$.}
\end{center}
\end{figure}

\begin{figure}
\includegraphics[width=0.95\columnwidth,angle=0]{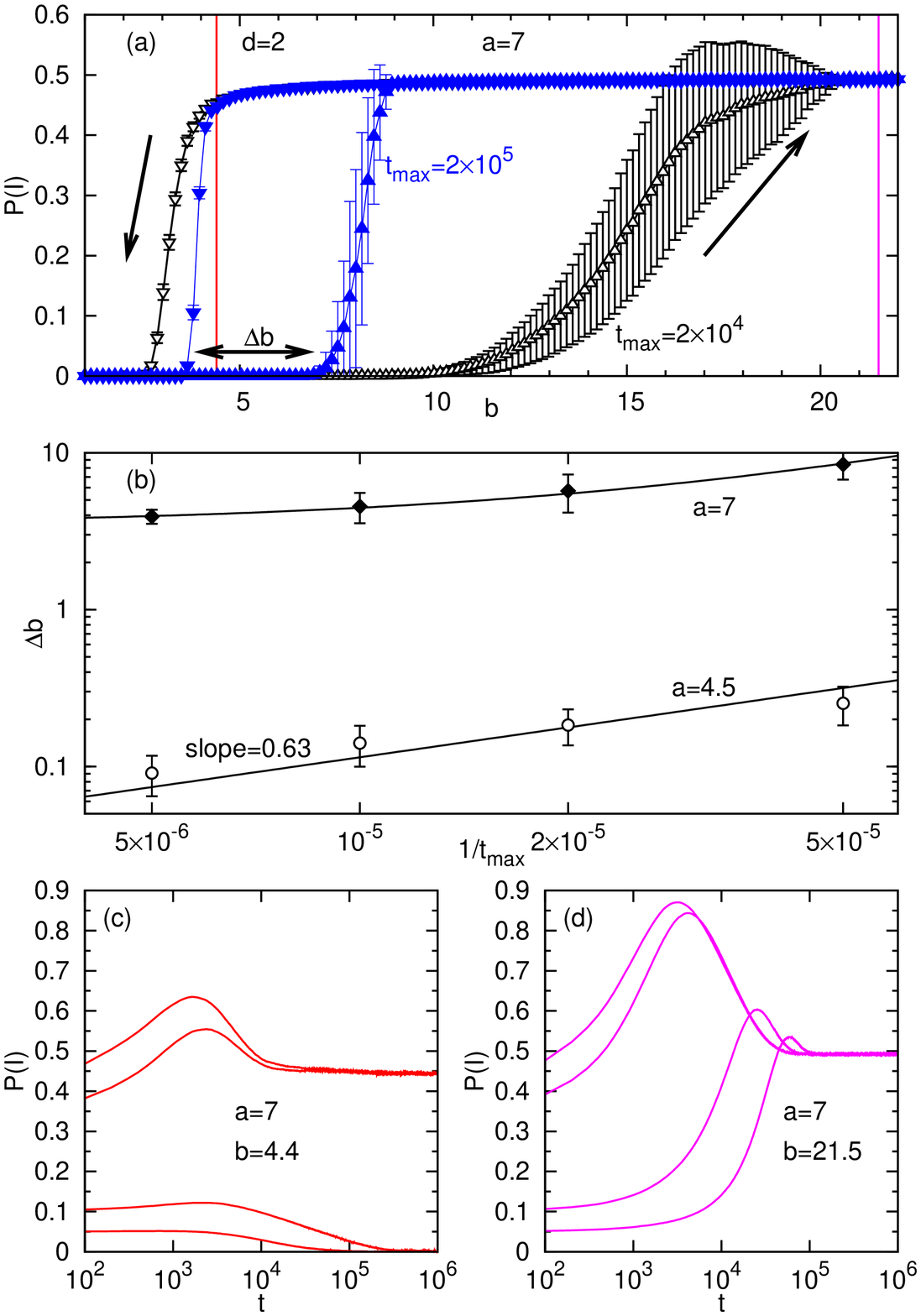}%
\begin{center}
\caption{\label{fig:2d}(Color online) Simulations for $d=2$ with
  $N=150^2$ and $h=2.5 \times 10^{-5}$. (a) ($a=7$, averaged over 10
  runs) The length of the hysteresis cycle $\Delta b$ decreases for
  increasing $t_{max}$ ($\Delta b$ denoted by the horizontal arrow for
  $t_{max}=2 \times 10^5$). Upward (downward) triangles denote
  increasing (decreasing) $b$. Vertical lines show values of $b$
  employed in panels (c) and (d). (b) Dependence of $\Delta b$
  on $t_{max}$ is qualitatively different for $a$ above (filled circles)
  and below (open circles) $a_c$, respectively, showing a nonzero or
  zero asymptotic value in the limit $t_{max}\to\infty$. The lower plot
  is well fitted by a power law with exponent 0.63. Note that the
  precision in $\Delta b$ is limited by the increment in $b$ employed
  in the hysteresis cycle ($\delta b=10^{-2}$ in the case $a=4.5$).
  Panels (c) and (d) show the time evolution of $P(I)$ (averaged over
  20 runs) for $a=7$ and different initial conditions [with $N P(I)$
  sites in state $I$ at $t=0$ and the remaining in state $S$], showing
  phase bistability for $b=4.4$ but not for $b=21.5$.}
\end{center}
\end{figure}

\begin{figure}
\includegraphics[width=0.95\columnwidth,angle=0]{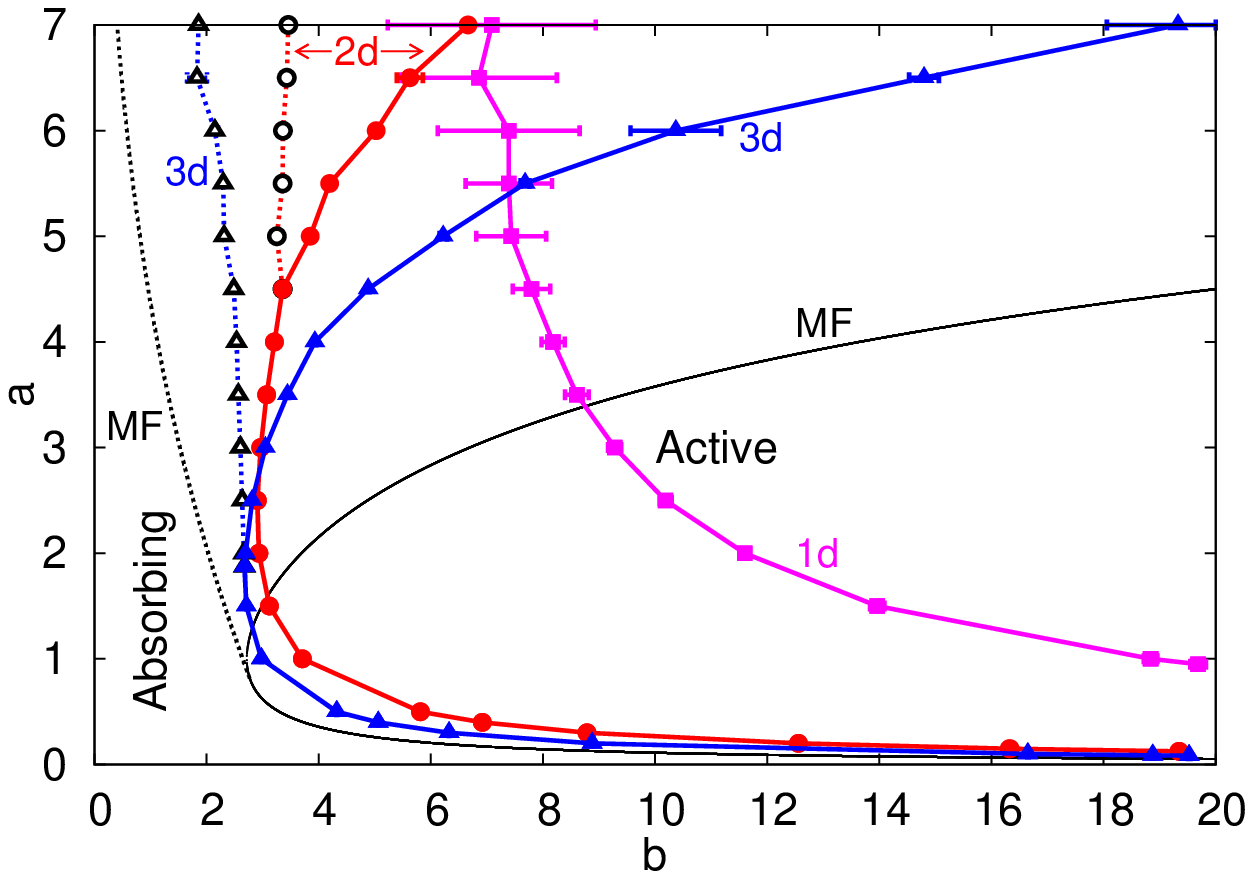}
\begin{center}
\caption{\label{fig:diagrama23d}(Color online) Phase diagram of the
  model in one-, two- and three-dimensional hypercubic lattices (squares,
  circles and triangles, respectively). Open symbols (with dashed
  lines to guide the eye) mark the onset of phase bistability, with
  the sudden appearance of a stable active phase with finite order
  parameter. Filled symbols (with solid lines to guide the eye) denote
  a transition in which the absorbing state loses stability.  The
  region in between dashed and solid lines correspond to the bistable
  regime. Lines without symbols show MF results for comparison. Results
  correspond to simulations with $\gamma=1$ and $h=2.5\times 10^{-5}$. 
  For $d=1$, 2, and 3, linear system sizes were $L=32400$, $L=150$, and $L=25$ 
  (where $N=L^d$), the maximum values of $t_{max}$ were $5\times 10^5$, 
  $2\times 10^5$, and $10^5$, with results averaged over 10, 10, and 20 runs, 
  respectively.}
\end{center}
\end{figure}

Since simulations with small values of $\gamma$ are very difficult to
perform~\cite{Joo04b}, we now focus on the simpler bifurcation
scenario of Fig.~\ref{fig:phasediagram} and discuss the results of
simulations for $\gamma=1$.  Identifying the nature of the transition
in hypercubic lattices is not so simple as for random and complete
graphs. As has been recently discussed in detail by
Takeuchi~\cite{Takeuchi08}, even a system which undergoes a continuous
phase transition into an absorbing state (such as those belonging to the
directed percolation universality class) may show a hysteresis cycle
when the coupling parameter loops around its critical value. This is
due to the divergence of the transient times at criticality in the
thermodynamic limit. In simulations, this gets reflected in the width
of the hysteresis cycle scaling with the ramp rate (defined as the
increment in $b$ per unit time) as $(1/t_{max})^{1/(\beta+1)}$, where
$\beta$ is the critical exponent governing the order
parameter~\cite{Takeuchi08,Marro99}.

Consider, for instance, the hysteresis cycles shown in
Fig.~\ref{fig:3d}(a) for $d=3$ and two values of $t_{max}$ differing
by an order of magnitude (while the increments in the values of $b$
have been kept the same~\cite{Takeuchi08}). Whether or not the
transition is continuous will depend on whether the width $\Delta b$
of the hysteresis cycle shrinks to zero in the limit $t_{max}\to
\infty$. We have operationally defined $\Delta b$ as follows: for each
run, we attribute the upper end of the hysteresis cycle to the $b$
value at which the density of active sites (averaged over $t_{max}$
time steps) $P(I)$ is first above $1/\sqrt{N}$. Similarly, the lower
end of the hysteresis cycle is defined as the point where $P(I)$ falls
below $1/\sqrt{N}$. The width of the hysteresis cycle $\Delta b$ is
then obtained by averaging the difference over the runs. We plotted
$\Delta b$ versus $1/t_{max}$ in Fig.~\ref{fig:3d}(b) for two values
of $a$. For $a=4.5$ [which corresponds to the hysteresis cycles shown
  in Fig.~\ref{fig:3d}(a)], a linear extrapolation leads to a nonzero
value of $\Delta b$ as $t_{max}\to \infty$, which is consistent with a
discontinuous phase transition. For $d=3$ and $a=1.875$, on the other
hand $\Delta b$ decreases to zero as a power law $(1/t_{max})^{0.55}$,
which is consistent with Takeuchi's prediction~\cite{Takeuchi08} for
the DP universality class [$\beta\simeq 0.805(10)$~\cite{Jensen92}].

One could in principle feel uncomfortable with the above described
criterion for deciding on the discontinuity of the transition because
in practice the linear extrapolation may not coincide with
$\lim_{t_{max}\to\infty}\Delta b$ if transient times are too long (and
we expect them to be long near a continuous transition!). This problem
becomes more salient as we decrease the spatial dimension, as depicted
for $d=2$ and $a=7$ in Fig.~\ref{fig:2d}. Note the smoothness of the
largest hysteresis cycle in Fig.~\ref{fig:2d}(a), which is very
similar to the ones observed by Takeuchi~\cite{Takeuchi08} near a {\em
  continuous\/} transition. According to the extrapolated value of
$\Delta b$ for $t_{max}\to \infty$, however, this transition would be
considered discontinuous [see upper plot in Fig.~\ref{fig:2d}(b)],
whereas for weaker nonlinearity [$a=4.5$, lower plot of
Fig.~\ref{fig:2d}(b)] we obtain again a power law
$(1/t_{max})^{0.63}$~\cite{Takeuchi08} compatible with DP in $d=2$
[$\beta\simeq 0.583(4)$~\cite{Jensen92}].

To be sure of the discontinuity of the transition, it is simpler (and
computationally less expensive) to investigate directly the alleged
bistability: we fix $a$ and $b$ and check the dependence of the
stationary state on the initial
condition~\cite{DickmanMaia08,OdorDickman09}. This test is shown in
Figs.~\ref{fig:3d}(c),~\ref{fig:3d}(d),~\ref{fig:2d}(c),~and~\ref{fig:2d}(d)
for $d=3$ and $d=2$, respectively. Figures~\ref{fig:3d}(c) and~\ref{fig:2d}(c)
confirm bistability since for lower (higher) initial values of $P(I)$ the
system converges to the absorbing (active) state. Figures~\ref{fig:3d}(d)
and~\ref{fig:2d}(d) serve as a negative control, confirming (in a region where
only the active state is stable) that the convergence to the absorbing state 
in the previous cases are not due to finite-size fluctuations. We note that all
samples converged to the same attractor (either absorbing or active) as their 
average.

The extrapolation $\lim_{t_{max}\to\infty}\Delta b$ was employed to
draw the phase diagram of the model for two- and three-dimensional
lattices. As shown in Fig.~\ref{fig:diagrama23d}, the qualitative
structure of the phase diagram is well reproduced by the mean-field
predictions, though quantitative agreement worsens as dimensionality
decreases as expected. Note that the bistable phase for $d=2$ is much
smaller than for $d=3$. For $d=1$, the large error bars in
Fig.~\ref{fig:diagrama23d} for large $a$ emerge due to extremely large
transients. We have not observed clearly discontinuous transitions for
$d=1$ up to $a=7$. This is in agreement with the results of 
Bidaux {\it et al.\/}~\cite{Bidaux89,Jensen91}, as well as with 
Hinrichsen's conjecture that discontinuous transitions in $d=1$ should 
only occur in the presence of diffusion~\cite{Hinrichsen00preprint}.

\section{\label{conclusions}Conclusions and perspectives}

In summary, we have proposed a Markovian continuous-time lattice model
of nonlinearly pulse-coupled excitable elements. Coupling depends
linearly on rate $b$ and nonlinearly on the dimensionless parameter
$a$. We have shown that increasing the nonlinearity of the coupling
leads to a change in the nature of the phase transition into an active
state.  In the regime of linear coupling ($a \ll 1$), where the model
approaches the stochastic SIRS model, an active phase with $P(I)>0$
appears through a continuous transition as $b$ increases. In a
sufficiently nonlinear regime (large enough $a$), increasing $b$ leads
to a discontinuous phase transition. The nature of the transition can
therefore be controlled by a single parameter, which is not
diffusion. These results can be predicted by mean-field analysis and
are qualitatively confirmed in simulations of random graphs and
hypercubic lattices for $d \geq 2$. The fact that a discontinuous
transition was not found for $d=1$ is consistent with previous results
for two-state systems with little or no
diffusion~\cite{Bidaux89,Jensen91, DickmanTome91,OdorBoccaraSzabo93,Hinrichsen00preprint}.

We have characterized discontinuous transitions by two complementary
criteria: first, hysteresis cycles were obtained and their width
estimated by extrapolation for an infinite number of Monte Carlo time
steps; then, bistability was explicitly confirmed by checking that the
system trajectory exhibits dependence on the initial conditions. In
the case of continuous transitions, the width of the hysteresis cycles
scaled with the ramp rate according to recent predictions by
Takeuchi~\cite{Takeuchi08}. 

Finally, we recall that the exponential coupling in Eq.~\ref{eq:ab}
was inspired in the model of Wood {\it et al.}~\cite{Wood06a}. While
their nonlinear phase-coupled stochastic oscillators do undergo a
phase transition into a synchronized state, we did not find sustained
collective oscillations with a similar nonlinearity among
pulse-coupled excitable elements. However, we did find (albeit in a
small parameter region) unstable global oscillations and collective
excitability in the mean-field equations. Simulations of the complete
graph revealed stochastic oscillations in single runs whenever the
active phase corresponded to a stable spiral in the mean-field
equations. It remains to be investigated whether collective
excitability and stochastic oscillations remain in regular lattices or
appear in the transition to a small-world regime.

\begin{acknowledgments}
V.R.V.A. and M.C. acknowledge financial support from CNPq, FACEPE, CAPES, and
special programs PRONEX and INCEMAQ. It is a pleasure to thank
R. Dickman for enlightening discussions during the preparation of this
work, as well as an anonymous referee for helpful comments on the
first version of the paper.
\end{acknowledgments}

\bibliography{copelli}

\end{document}